\documentclass[namedreferences]{SolarPhysics}
\usepackage{solaheader}    
\usepackage[optionalrh]{spr-sola-addons} 
\usepackage{graphicx}        
\usepackage{color}           
\usepackage{url}             

\newcommand{\solareceiveddate}{11 May 2009} 
\newcommand{\solaaccepteddate}{18 August 2009}                                  

\newcommand{\etal}{{\it et al.}}

\newcommand{\aap}{    {\it Astron. Astrophys.}}

\newcommand{\apj}{    {\it Astrophys. J.}}

\newcommand{\solphys}{{\it Solar Phys.}}

\newcommand{\ssr}{    {\it Space Sci. Rev.}}

\begin{document}

\begin{article}

\begin{opening}

\title{Spectro-Polarimetric Observations of Solar Magnetic Fields and the SOHO/MDI Calibration Issue}

\author{M.L.~\surname{Demidov}$^{1}$\sep H.~\surname{Balthasar}$^{2}$}
\runningauthor{M.L.Demidov, H.Balthasar} 
\runningtitle{Solar Magnetic fields and the SOHO/MDI Calibration}

   \institute{$^{1}$ Institute of Solar-Terrestrial Physics, Siberian
   Branch, Russian Academy of Sciences, 664033 Irkutsk,  P.O. Box 291,   Russia\\
   email: \url{demid@iszf.irk.ru} \\
     $^{2}$ Astrophysikalisches Institut Potsdam (AIP), An der Sternwarte 16,
             14482 Potsdam, Germany \\
                    email: \url{hbalthasar@aip.de} \\}

\begin{abstract} 
Comparisons of solar magnetic-field measurements made in different spectral lines are very important, especially in those lines in which observations have a long history or(and) specific diagnostic significance. The spectral lines 
Fe\,{\sc i}~523.3\,nm  and Fe\,{\sc i}~525.0\,nm belong to this class. Therefore, this study is devoted to a comprehensive analysis using new high-precision Stokes-meter full-disk observations. The  disk-averaged magnetic-field strength 
ratio $R = B(523.3)/B(525.0)$ equals $1.97\,\pm 0.02$. The center-to-limb variation (CLV) is  $R = 1.74 - 2.43\mu + 3.43\mu^{2}$, where $\mu$ is the cosine of the
center-to-limb angle. For the disk center, we find $R = 2.74$, and for near-limb areas with 
$\mu = 0.3$, $R$ equals 1.32. There is only a small dependence of $R$ on the spatial resolution. Our results are rather close to those 
published three decades ago, but differ significantly from recent magnetographic 
observations.
An application of our results to the important SOHO/MDI magnetic data calibration issue is discussed.
We conclude 
that the revision of the SOHO/MDI data, based only on the comparison of magnetic field measurements in the line pair
Fe\,{\sc i}~523.3\,nm and Fe\,{\sc i}~525.0\,nm (increasing  by a factor of 1.7 or 1.6 on average according to recent publications) is not obvious and new investigations are urgently 
needed.
\end{abstract}
\end{opening}

\section{Introduction}
     \label{S-Introduction}

Many tasks of solar and solar-terrestrial physics urgently require
precise, quantitative information about the distribution of magnetic fields across the solar disk. 
However, the extremely complicated spatial structure of magnetohydrodynamic parameters in the solar atmosphere leads to 
differences in magnetic field data from different spectral lines.
This is the reason why comprehensive analyses of magnetic-field measurements made in different spectral lines are important for 
diagnostics of solar magnetism and for the analysis of different data 
sets. In the first case, it is better to use observations performed at the same instrument, to avoid possible influences of different instrumental, method, time-dependent, and other effects. In the second case, the 
aim is to construct composite long-term data sets combining
different observatories to study, for example, the temporal variations of solar magnetic parameters on different time scales.

A very significant part of solar magnetic-field observations has been performed (and are performed up to now) in the spectral line Fe\,{\sc i}~525.0\,nm. Therefore, comparisons  of observations in other spectral lines with observations in this particular line are extremely important. The atomic parameters of Fe\,{\sc i}~525.0\,nm 
(large  Land\'e factor, $g$ = 3, small lower level excitation potential, $\chi_{\it l}$ = 0.12 eV) stimulated vigorous 
discussions about the reliability of measurements in this line, which are not finished yet. Different authors (some aspects of this issue are given in the discussion in \opencite{Demidov-etal08} and in the recent paper of \opencite{Ulrichetal09}) offer different correction factors to reduce raw measurements in 
Fe\,{\sc i}~525.0\,nm to the ``true'' ones. 

Because of its low temperature sensitivity, one of the lines used as a ``standard'' for such correction is the rather strong line 
Fe\,{\sc i}~523.3\,nm  with $g_{\rm eff} = 1.3$ and $\chi_{\it l}$ = 2.93 eV  (\opencite{HarLiv69}; \opencite{HSt72}; \opencite{FrSt72}; \opencite{FrSt78}; \opencite{Ulrich92}; \opencite{Ul-etal02}; \opencite{Tran-etal05}; \opencite{Ulrichetal09}). 
However, the combination of this line and Fe\,{\sc i}~525.0\,nm is far from being perfect for diagnostics of solar magnetic fields, because they have different thermal dependencies, and Fe\,{\sc i}~523.3\,nm is not intrinsically narrow (see \opencite{Solanki93}).
With much more success, it could be used as an example for the thermal line ratio. As shown by \inlinecite{Socas-etal08}, reliable information about the magnetic-field parameters in the quiet regions (which cover most of the solar surface) can be achieved only using a combination of several lines with particular characteristics. 

Nevertheless,  observations in the discussed lines are still important in the context of the cross-calibration of different data sets, especially for the data recalibration problem of the
{\it Michelson Doppler Imager} (MDI) onboard the {\it Solar and Heliospheric Observatory} (SOHO), see {\it http://soi.stanford.edu/magnetic} and Solar News (issues No. 19, 2007 and No. 25, 2008). The SOHO/MDI observations of magnetic fields are performed in the spectral line
Ni\,{\sc i}~676.8\,nm with $g_{\rm eff} = 1.43$. The main reasons for two subsequent serious revisions of the magnetic field data from MDI, 
which are widely used in the scientific community, are based on the comparison of observations in Fe\,{\sc i}~523.3\,nm and Fe\,{\sc i}~ 525.0\,nm made at the Mount Wilson Observatory (MWO) and published by \inlinecite{Tran-etal05} and \inlinecite{Ulrichetal09}. According to the first paper, based on the results of \inlinecite{Ulrich92} and \inlinecite{WangSheeley95},  observations in Fe\,{\sc i}~525.0\,nm must be multiplied by the factor 
$\delta^{-1} =  B(523.3)/B(525.0) = 4.5 -2.5\sin^{2}\theta$
or, according to the second paper, by the factor $\delta^{-1} = 4.15 -2.82\sin^{2}\theta$, where $\theta$ is the center-to-limb angle. As a consequence, to fit SOHO/MDI data to MWO observations in 
Fe\,{\sc i}~525.0\,nm  corrected in this way, they must be multiplied by a factor that is a function of $\theta$, see Figure 6 and Table 3 of \inlinecite{Tran-etal05} and whose mean value over the disk is 1.7 . According to \inlinecite{Ulrichetal09}, the factor is 1/0.619 = 1.615 for $\theta \leq 30^{\circ}$.

The fact, that the new MWO results strongly contradict  the previous ones of \inlinecite{HSt72}, who found 
that the center-to-limb variation (CLV) of the coefficient is 
$\delta^{-1} = 0.48 + 1.33\mu$, with $\mu = \cos\theta$, requires independent measurements.
For  disk center, the factor is only 1.8 instead of 4.12 according to \inlinecite{Ulrichetal09}. 
Their factor between $B(523.3)$ and $B(525.0)$ can be up to 5.5 (see their Figure 3), 
which can hardly be explained by ``the fact that the shifted Zeeman components at $\lambda 525.0$\,nm are shifted beyond the sampling passband of the MWO Babcock magnetograph'' (\opencite{Ulrichetal09}).  

The above mentioned circumstances make an additional exploration of solar magnetic field observations in 
Fe\,{\sc i}~523.3\,nm and Fe\,{\sc i}~525.0\,nm an urgent task. 
The entire spectral-line profile should be considered.
This is the main aim of our study.

\section{Observations and Analysis} 
      \label{S-Observations}

All previous observations concerning the relationship between magnetic-field data in Fe\,{\sc i}~523.3\,nm and Fe\,{\sc i}~525.0\,nm were obtained with photoelectrical magnetographs. Drawbacks of such data are evident, because 
direct information about the distribution across the spectral line profile of the Stokes parameters (Stokes $V/I_{c}$ in the case of longitudinal field) is missing -- the only reliable indicator of magnetic field in the spatial range of line formation. Some assumptions and calibrations are necessary to obtain magnetic-field strengths from intensity variations in the exit slits of the magnetograph, which might be different for different spectral lines in the case of simultaneous observations. 

Spectropolarimetric observations, as used in the present study,  are much more reliable and informative. To study the problem of the comparison of solar magnetic-field measurements in combination of these lines, several series of observations in different regimes 
have been performed with the Stokes-meter of the {\it Solar Telescope for Operative Predictions} (STOP) at the Sayan Solar Observatory (SSO). Basic information about this instrument and methodical issues can be found in \inlinecite{Demidov-etal02} and \inlinecite{Demidov-etal08}. STOP is 
equipped with a linear, 29 mm wide Toshiba TCD CCD detector with 3424 pixels (height 200 $\mu$m, width 8 $\mu$m), which allows us to obtain high-precision measurements of the Stokes 
parameters  $I$ and $V/I_{c}$ in several spectral lines simultaneously.
The solar surface is scanned following a 2D-raster, and the spatial resolution depends
on defocussing the solar image.
The standard angular resolution  used in the regular observing programs (magnetograms of the large-scale magnetic fields, LSMF) is $100^{\prime\prime}$ and
the scanning step is $91^{\prime\prime}$. 
To cover both lines with a single exposure, we used the fourth order of the spectrograph, where more than 2.7 nm are available. Part of the observations obtained with this dispersion was done with the standard  spatial resolution ( Figures 7 and 8), but most
with focused solar image and standard scanning step, what corresponds to a resolution of $\approx 10^{\prime\prime}$.

The cycle time (one phase of the electro-optical modulator -- $KD^{*}P$ crystal in our case) was 30\,ms, the signal integration time in every point of the scanning process across the Sun was four seconds with and four seconds without the half-wave plate.
The $\lambda/2$ plate is periodically inserted into the light beam in front of the coelostat to monitor the zero-level position.

\begin{figure}    
   \centerline{\hspace*{0.015\textwidth}
             \includegraphics[width=0.90\textwidth,clip=]{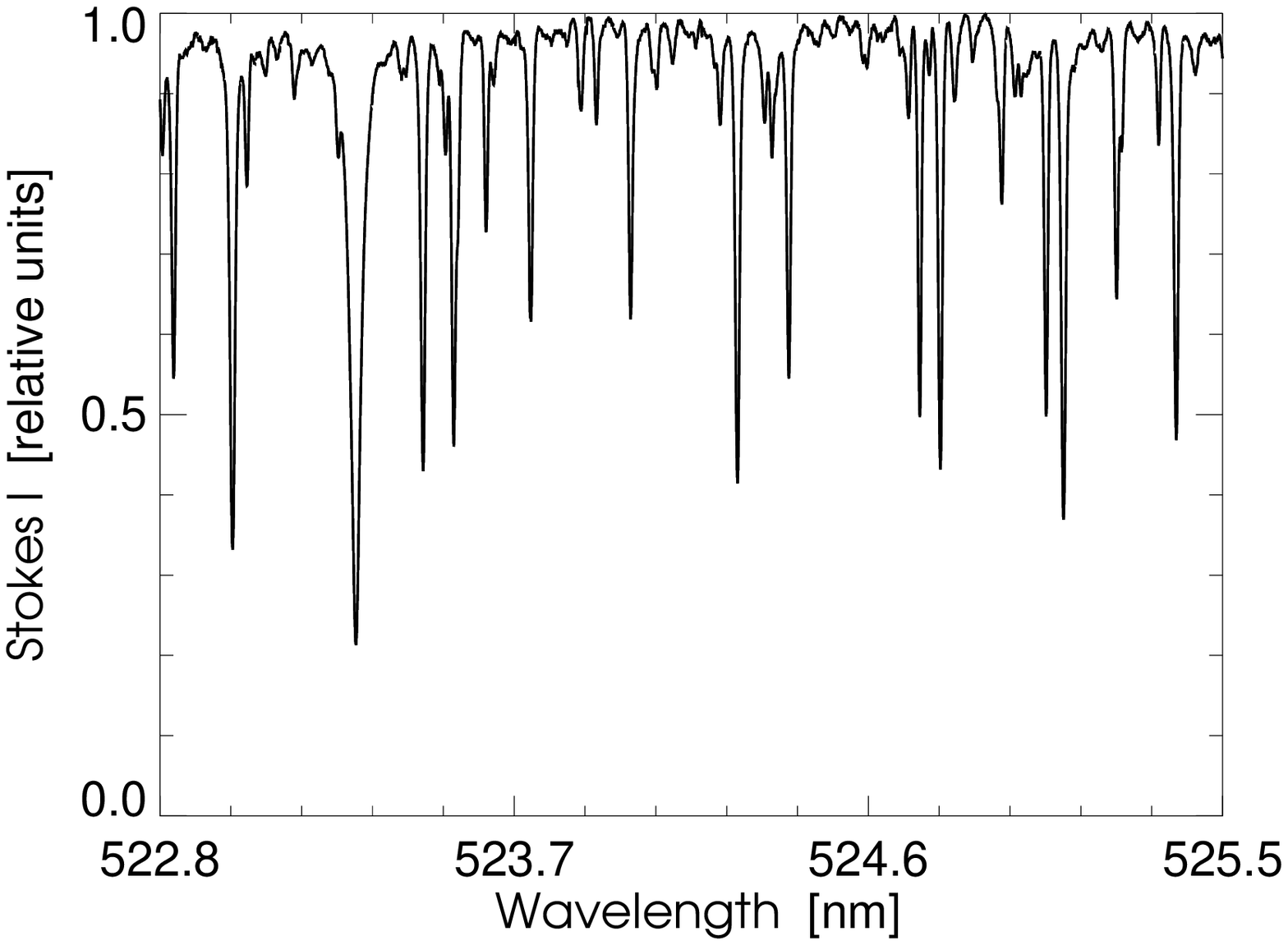}
                }
     
     \vspace{0.05\textwidth}    
   \centerline{\hspace*{0.015\textwidth}
           \includegraphics[width=0.92\textwidth,clip=]{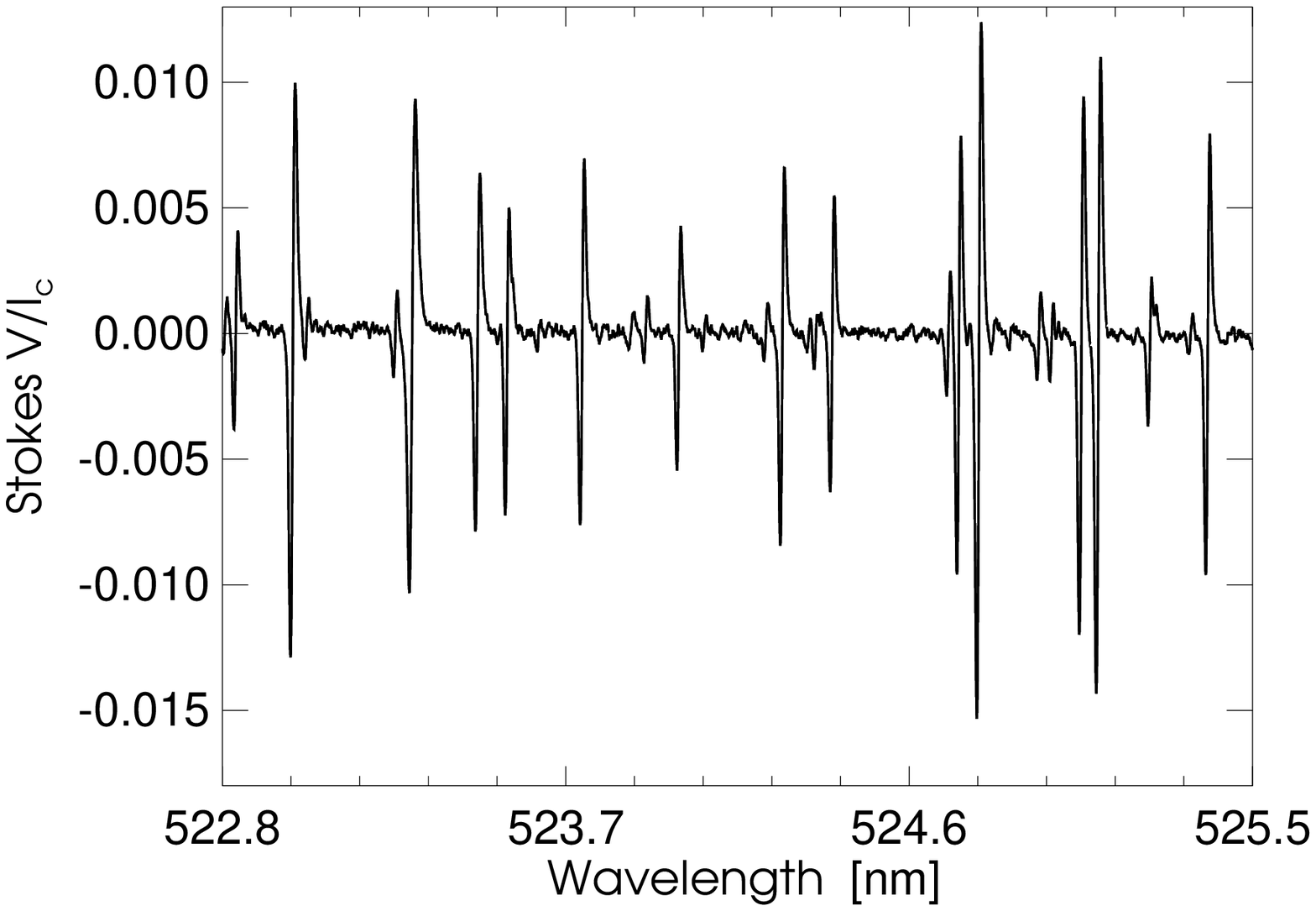}
}
 
     \vspace{0.1\textwidth}    

 \caption{Spectra of Stokes parameters $I$ (top panel) and $ V/I_{c}$
 (bottom panel) along the whole wavelength band including 
 Fe\,{\sc i}~523.3\,nm and Fe\,{\sc i}~525.0\,nm, covered by 
 the detector.
 These observations were obtained close to the solar disk
 center with a magnetic-field strength of --26\,G measured in 
 Fe\,{\sc i}~525.0\,nm. 
        }
   \label{F-I-V-all}
   \end{figure}

An example of Stokes $I$ and Stokes $V/I_{c}$ profiles across the whole CCD detector is shown in 
Figure \ref{F-I-V-all}. The data correspond to the point next to the solar disk center, 
observed on 3 February, 2009 with a resolution of about  $10^{\prime\prime}$. It is easy to see how much information is provided by such observations. 

To derive   quantitative information about magnetic and thermodynamic parameters from Stokes profiles such as those shown in Figure \ref{F-I-V-all} is a complicated task. We use the  approach (see \opencite{Demidov-etal02}), which numerically  imitates the work of solar magnetographs in the center-of-gravity mode (see \opencite{Solanki93}), to obtain the magnetic flux density (in the following we will use the term  ``magnetic field strength'' for it throughout the paper) from the recorded data.
The code calculates the position of the spectral line in two states ($\pm\lambda$/4) of the polarization analyzer. 
Parameters for the ``exit slits'' are selected corresponding to the  different lines. This position corresponds to the middle between the ``slits'', where the intensities are equal. If $\lambda_{1}$ is the wavelength of the spectral line in one analyser state and $\lambda_{2}$ in the other one, the value $\Delta \lambda = (\lambda_{1} - \lambda_{2})$/2 is considered as the Zeeman shift associated with the magnetic-field strength $B$ by the known equation $\Delta \lambda  = 4.67\times 10^{-5} g\lambda^{2}B$, where $B$ is in Gauss and 
$\lambda$ in cm.  For the case shown in Figure \ref{F-I-V-all}, the magnetic strength calculated this way is --26\,G for Fe\,{\sc i}~525.0\,nm and --64 G for Fe\,{\sc i} 523.3\,nm.
Parameters for the ``exit slits'' are: the width $W$ is equal to 9.32 pm, the separation $\Delta$ (distance between line center and the ``slits'' centers) is equal to 6.21 pm for the first line and $W = 24.84$ pm, $\Delta = 15.53$ pm for the second line. 
\begin{figure}    
   \centerline{\hspace*{0.015\textwidth}
               \includegraphics[width=0.90\textwidth,clip=]{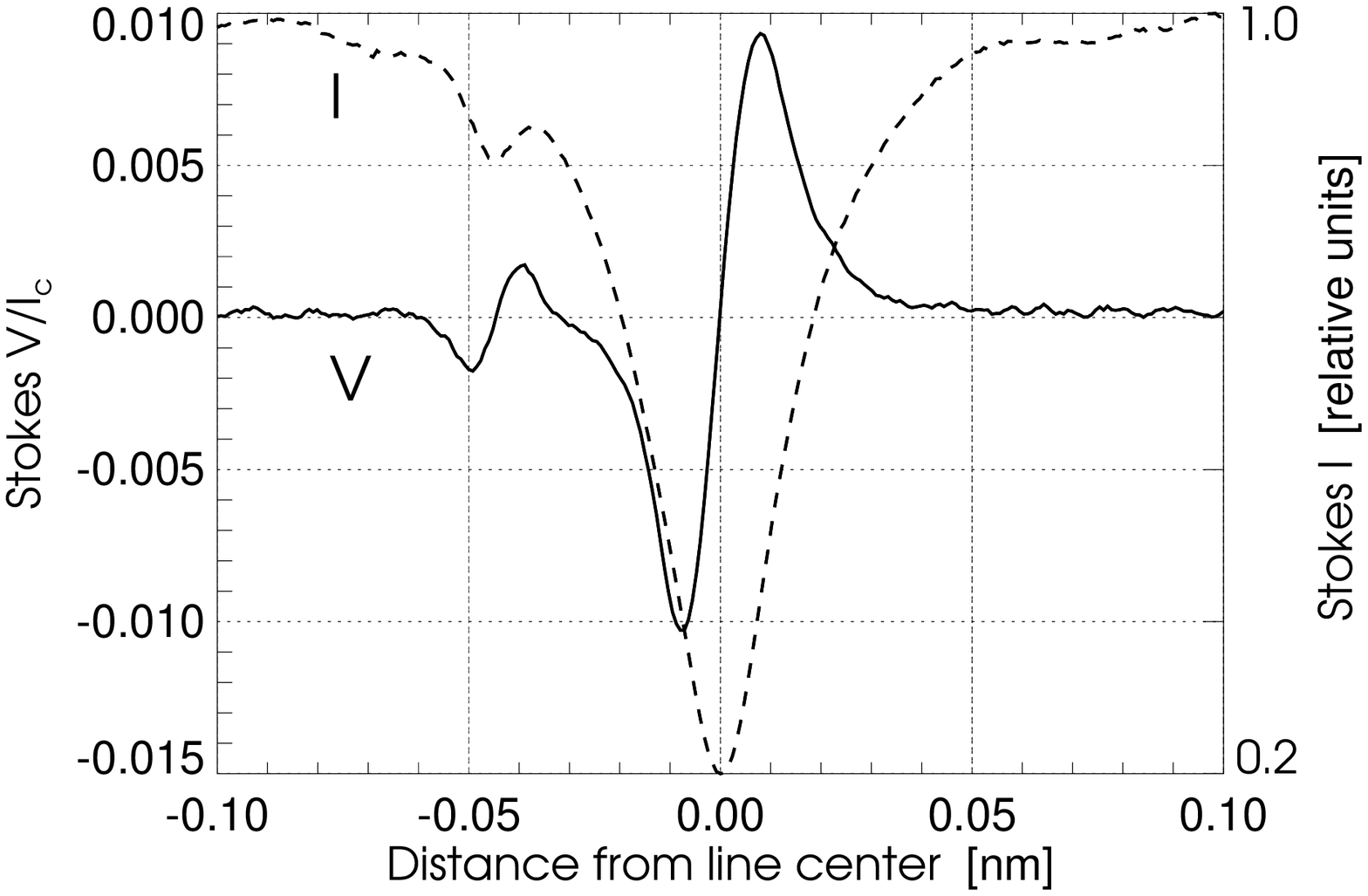}
                             }
          \vspace{0.05\textwidth}    
   \centerline{\hspace*{0.015\textwidth}
               \includegraphics[width=0.92\textwidth,clip=]{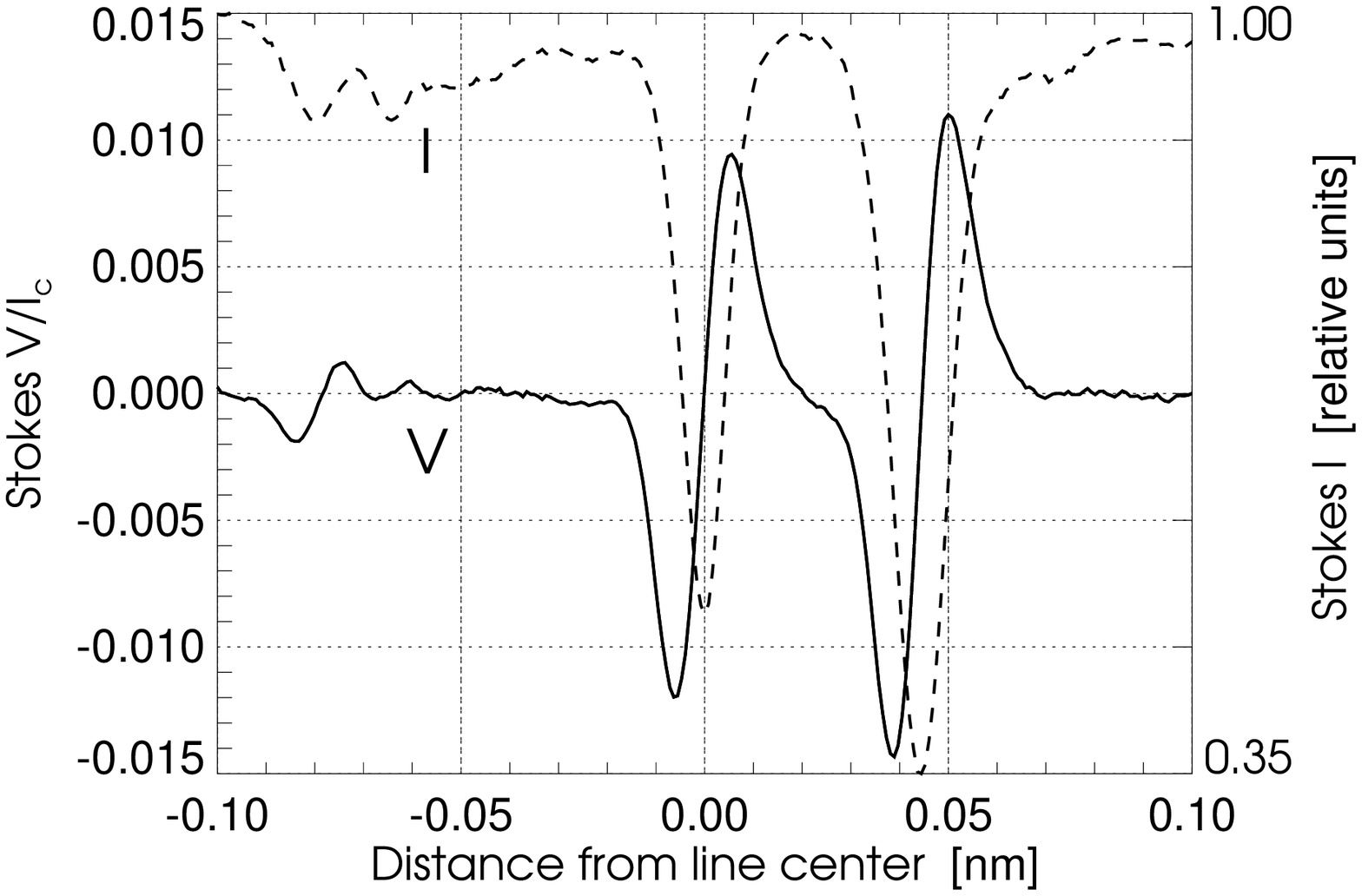}
                            }
          \vspace{0.1\textwidth}    

 \caption{Examples of Stokes $I$ (in relative units and normalized 
to fit the scale of graph) and Stokes $V/I_{c}$ profiles for 
Fe\,{\sc i}~523.3\,nm (top panel) and for Fe\,{\sc i}~525.0\,nm (bottom panel). The same observations as in Figure \ref{F-I-V-all} were used.
        }
   \label{F-I-V-U-M}
   \end{figure}
\begin{figure}    
   \centerline{\hspace*{0.015\textwidth}
               \includegraphics[width=0.90\textwidth,clip=]{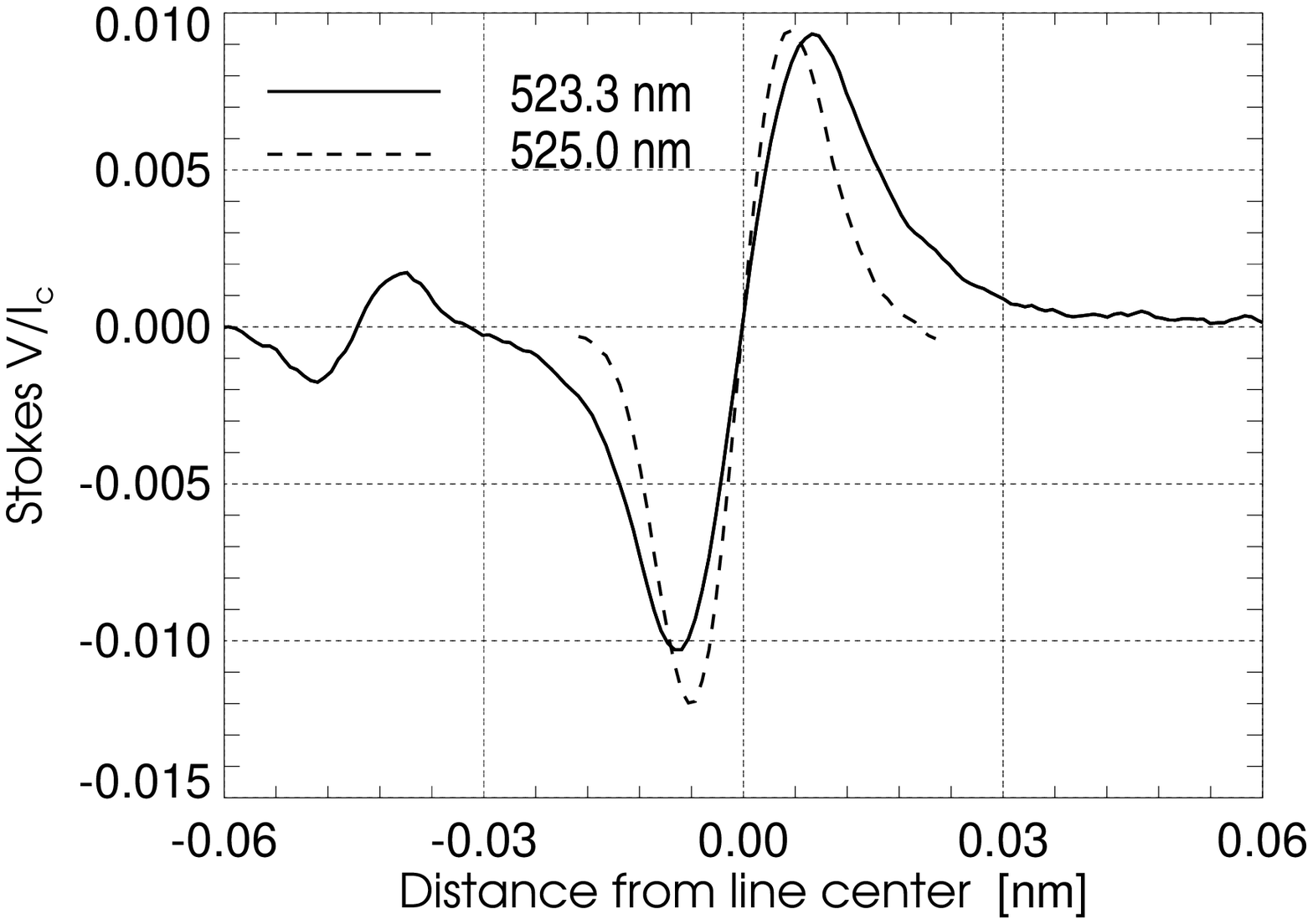}
                             }
     
     \vspace{0.01\textwidth}    

 \caption{Example of Stokes $V/I_{c}$ profiles of Fe\,{\sc i}~523.3\,nm and 
Fe\,{\sc i}~525.0\,nm. The same observations as in Figure \ref{F-I-V-all} and Figure \ref{F-I-V-U-M} were used.
        }
   \label{F-V-U-M}
   \end{figure}

The data used for  Figure \ref{F-I-V-all} offer the possibility to estimate the errors of our measurements. To do this the noise level of the Stokes $V/I_{c}$  signal in the continuum can be used. The best choice  for it is the part of the spectrum on the right of Fe\,{\sc i}~524.4\,nm. Calculations made for the $\approx$ 150\,pm wide band  show that the rms value in this region is $1.2\times 10^{-4}$. Taking into account that $V/I_{c} = $0.01 corresponds to $\approx$ 26\,G for Fe\,{\sc i}~525.0\,nm  and  $\approx$ 64\,G for Fe\,{\sc i} 523.3\,nm, we conclude that the formal error of the measurements is $\approx$ 0.3\,G for the first line and $\approx$ 0.8\,G for the second.

  Figure \ref{F-I-V-U-M} shows the Stokes $I$ and Stokes $V/I_{c}$ profiles for this pair of lines 
from the same observations as in Figure \ref{F-I-V-all}. Figure \ref{F-V-U-M} 
compares the Stokes $V/I_{c}$ profiles of both lines. 
Despite of the large difference in the $g$-factors, the amplitudes of the Stokes $V/I_{c}$-profiles are almost the same and   the peak positions of the Stokes $V/I_{c}$ profiles are close to the steepest parts of the corresponding Stokes $I$ profiles. As a consequence of the fact that Fe\,{\sc i}~523.3\,nm is deeper and 
wider than Fe\,{\sc i}~525.0\,nm, the separation of the Stokes $V/I_{c}$ extrema for the first 
line is almost  two times larger than for the second one, although the splitting 
factor of the latter is larger. Therefore, in the case of weak fields, the separation of the Stokes $V/I_{c}$ profiles peaks cannot be a measure of the magnetic field strength when lines with different widths are used. Indeed, if the magnetic-field strength is far from the  Zeeman saturation regime, the locations of the Stokes $V/I_{c}$ profile peaks are  insensitive to the field and determined from the  derivative of the Stokes $I$ profile.

Already Figure \ref{F-V-U-M} allows for the disk center a preliminary estimate of the ratio
$R = B(523.3)/B(525.0)$.
In agreement with \inlinecite{Demidov-etal08}, we prefer the
notation $R$ for the ratio or the regression coefficient.
The ratio of the Stokes $V/I_{c}$ amplitudes (average of the amplitudes of the blue and red Stokes $V/I_{c}$ peaks) of the two lines is $V(523.3)/V(525.0) = 0.0097/0.0106 = 0.915$. Normalizing to the corresponding Land\'e factors, we obtain $R = 0.915\times(3.0/1.3) = 0.915\times2.3 = 2.1$. This value is rather close to the estimation $R = B(523.3)/B(525.0) = (-64\,{\rm G}/-26\,{\rm G}) = 2.4$, obtained from calculations of the magnetic-field strengths using wide parts of the line wings.  

\begin{figure}    
   \centerline{\hspace*{0.015\textwidth}
               \includegraphics[width=0.90\textwidth,clip=]{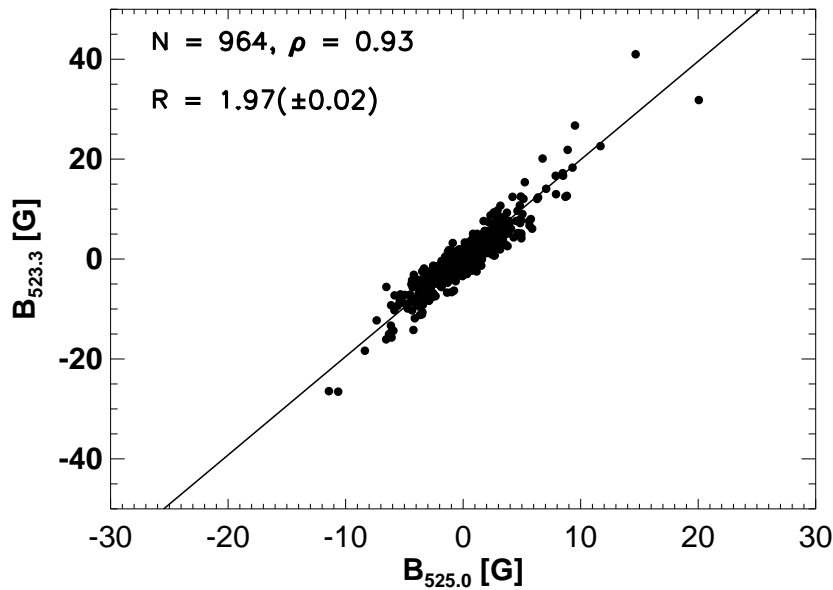}
                             }

     \vspace{0.01\textwidth}    

 \caption{Correlation and regression analysis of solar magnetic field observations in 
Fe\,{\sc i}~523.3\,nm and Fe\,{\sc i}~525.0\,nm. Data for three days of observations, 1, 2 and 3 February, 2009, are used. $N$ is the number of points, $\rho$ is the correlation coefficient, and $R$ is the coefficient of linear regression (the slope of the line through the scatter plot). 
        }
   \label{F-scatterplot-U-M}
   \end{figure}

One point is not enough for a reliable estimation of the 
$B(523.3)/B(525.0)$ ratio. The scatter plot, calculated for three full-disk 
magnetograms observed on 1, 2, and 3 February, 2009, is shown in Figure \ref{F-scatterplot-U-M}.
Parameters for the ``exit slits'' are the same as  mentioned above.  We  see a rather high correlation coefficient 
$\rho = 0.93$, and a systematic difference with regression coefficient
$R = 1.97 \pm0.02$. To calculate $R$, the reduced major axis method (\opencite{Davis86}) is used. A formula to  estimate the error (which is determined from  the scattering of the points, and which is less when the correlation coefficient is higher) is taken from this book too.  According to our measurements, the average factor that should be used to adjust 
measurements in Fe\,{\sc i}~525.0\,nm to those in 
Fe\,{\sc i}~523.3\,nm, is $1.97 \pm0.02$.

\begin{figure}    
   \centerline{\hspace*{0.015\textwidth}
               \includegraphics[width=0.90\textwidth,clip=]{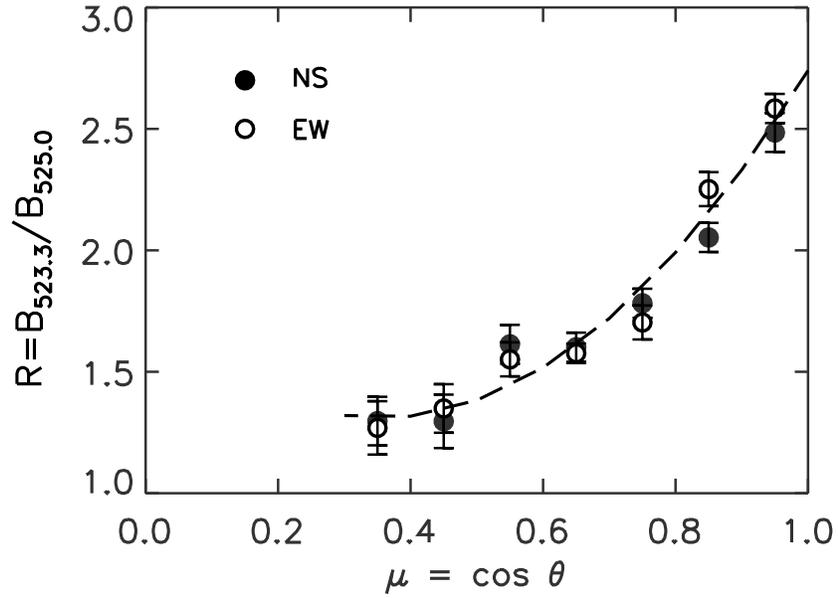}
                             }
     
     \vspace{0.05\textwidth}    

 \caption{Center-to-limb variation of the ratio  
$R = B(523.3)/B(525.0)$ for the polar (N\,--\,S) and equatorial (E\,--\,W) sectors of the solar disk.}   
   \label{F-CLV-R}
   \end{figure}

The next logical step is to check for possible spatial variations of the ratio $R = B(523.3)/B(525.0)$  across the solar disk. 
For this purpose, we divide the solar disk into polar and equatorial sectors as in \inlinecite {Demidov-etal08}.
We derive a polynomial fit $ R = 1.74 - 2.43\,\mu + 3.43\,\mu^2$, shown in
Figure \ref{F-CLV-R}. There are no significant differences in the CLV of $R$ between polar and equatorial sectors of the disk. At disk center, $R$ is 2.74, 
and closer to the limb, at $\mu$ = 0.3, $R$ is 1.32. The dashed line is a quadratic polynomial fit through the observed points.

\begin{figure}    
   
   \centerline{\hspace*{0.015\textwidth}
              \includegraphics[width=0.90\textwidth,clip=]{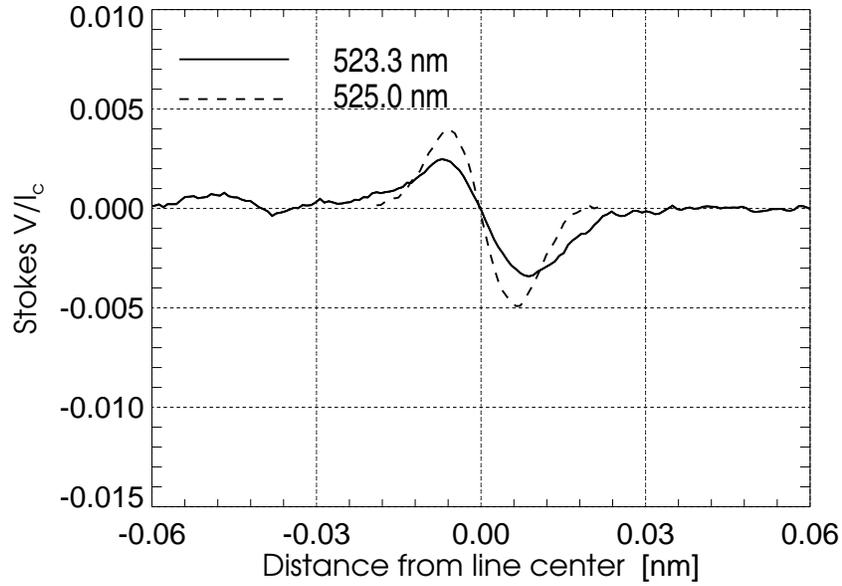}     
                     }
     
     \vspace{0.05\textwidth}    

 \caption{Example of Stokes $V/I_{c}$ profiles of Fe\,{\sc i}~523.3\,nm and 
Fe\,{\sc i}~525.0\,nm  for a point near the south pole with $\mu = 0.56$ and with a magnetic-field
 strength $B$ = 11.6\,G from Fe\,{\sc i}~525.0\,nm.}
   \label{F-V-U-M-SOUTH}
   \end{figure}

A significant decrease of the ratio $B(523.3)/B(525.0)$
with increasing heliocentric distance is illustrated in 
Figure \ref{F-V-U-M-SOUTH}, where $V/I_{c}$ profiles of the analyzed lines are shown for the point next to the south pole with $\mu = 0.56$ and with magnetic field strengths $B(525.0) = 11.6\,$G and 
$B(523.3) = 19.8\,$G. The ratio of these strengths, $R\approx1.5$, is rather close to $R = 1.45$ from the analytical formula in Figure \ref{F-CLV-R} at $\mu = 0.56$.
 
As mentioned above, observations in   Fe\,{\sc i}~523.3\,nm and Fe\,{\sc i}~525.0\,nm with the STOP traditional spatial resolution of 
$100^{\prime\prime}$ were performed besides the observations with the rather high spatial resolution in order to look for a possible dependence of $R = B(523.3)/B(525.0)$ on the spatial resolution. \inlinecite {Ulrich92} found a significant increase of $R$ with decreasing spatial resolution: $R$ = 4.5 at disk center at a resolution of $20^{\prime\prime}$  instead of $R = 3.9$ at a resolution of $5^{\prime\prime}$. 

\begin{figure}    
   \centerline{\hspace*{0.015\textwidth}
               \includegraphics[width=0.48\textwidth,clip=]{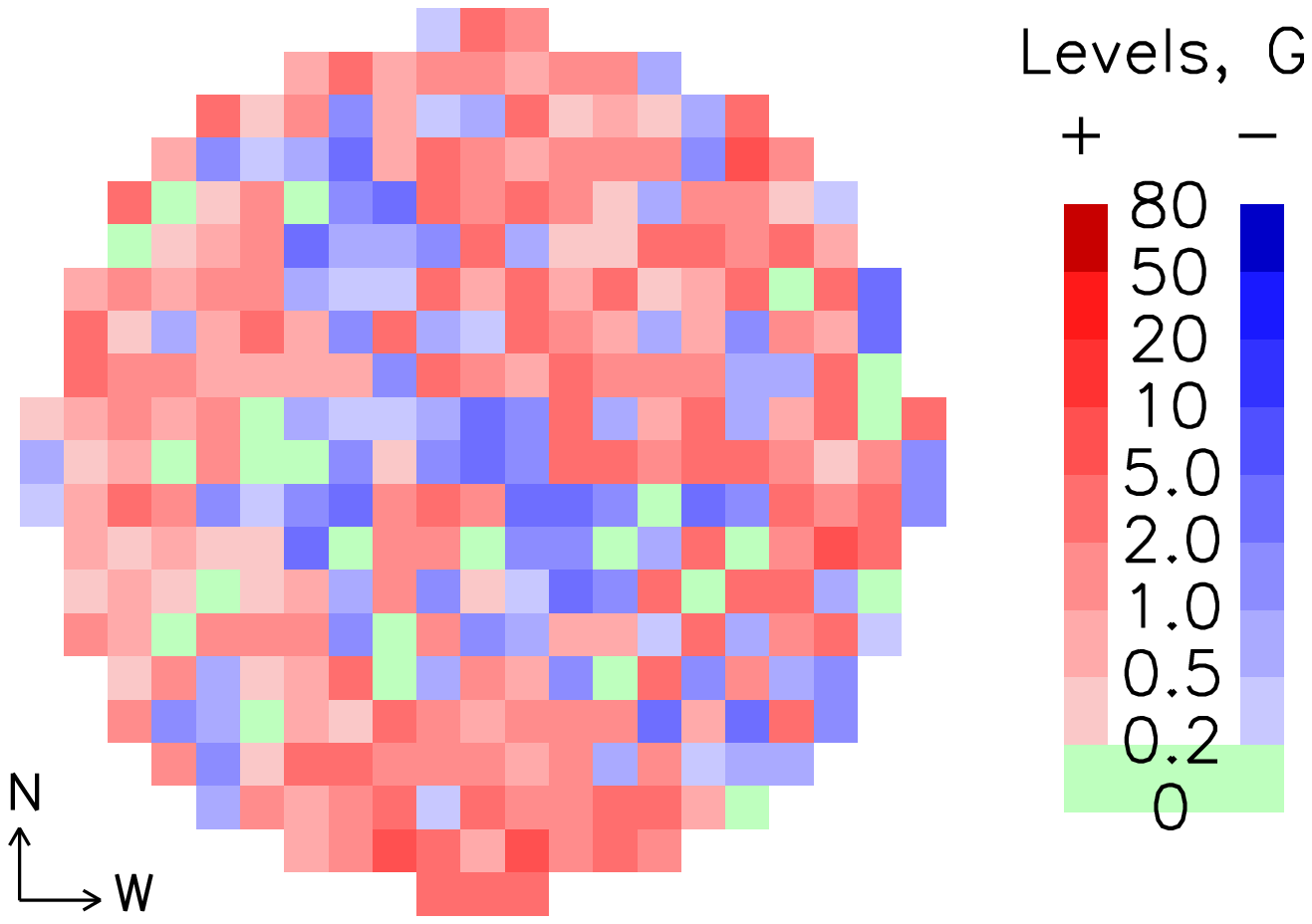}
              \hspace*{0.02\textwidth}
               \includegraphics[width=0.48\textwidth,clip=]{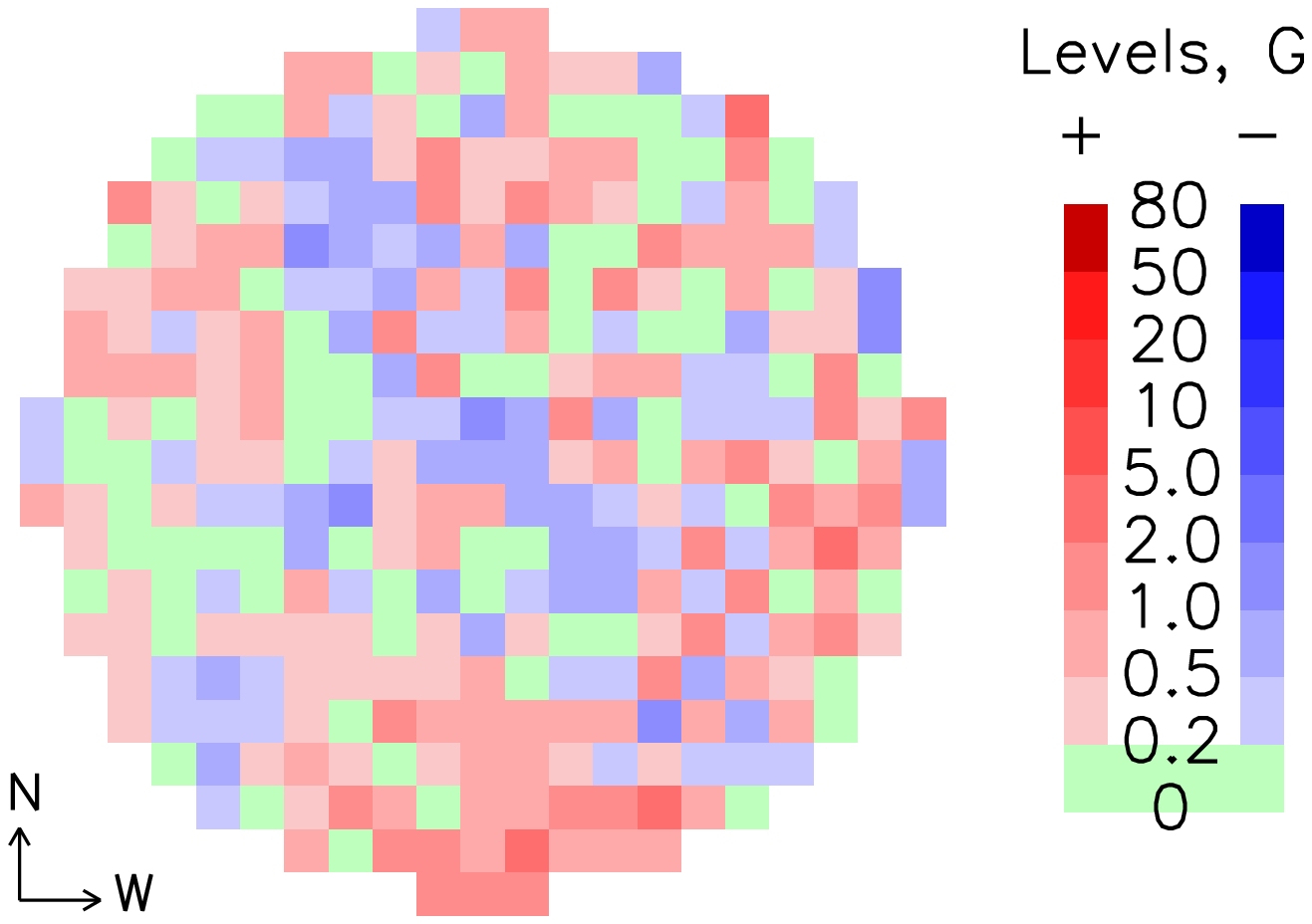}
              }
 
     \vspace{0.05\textwidth}    

 \caption{Full-disk magnetograms observed with a spatial resolution of 
$100^{\prime\prime}$ in Fe\,{\sc i}~523.3\,nm (left panel) and Fe\,{\sc i}~525.0\,nm (right panel) on  1 February, 2009.
}   
   \label{F-2magnetograms}
   \end{figure}
\begin{figure}    
   \centerline{\hspace*{0.015\textwidth}
               \includegraphics[width=0.90\textwidth,clip=]{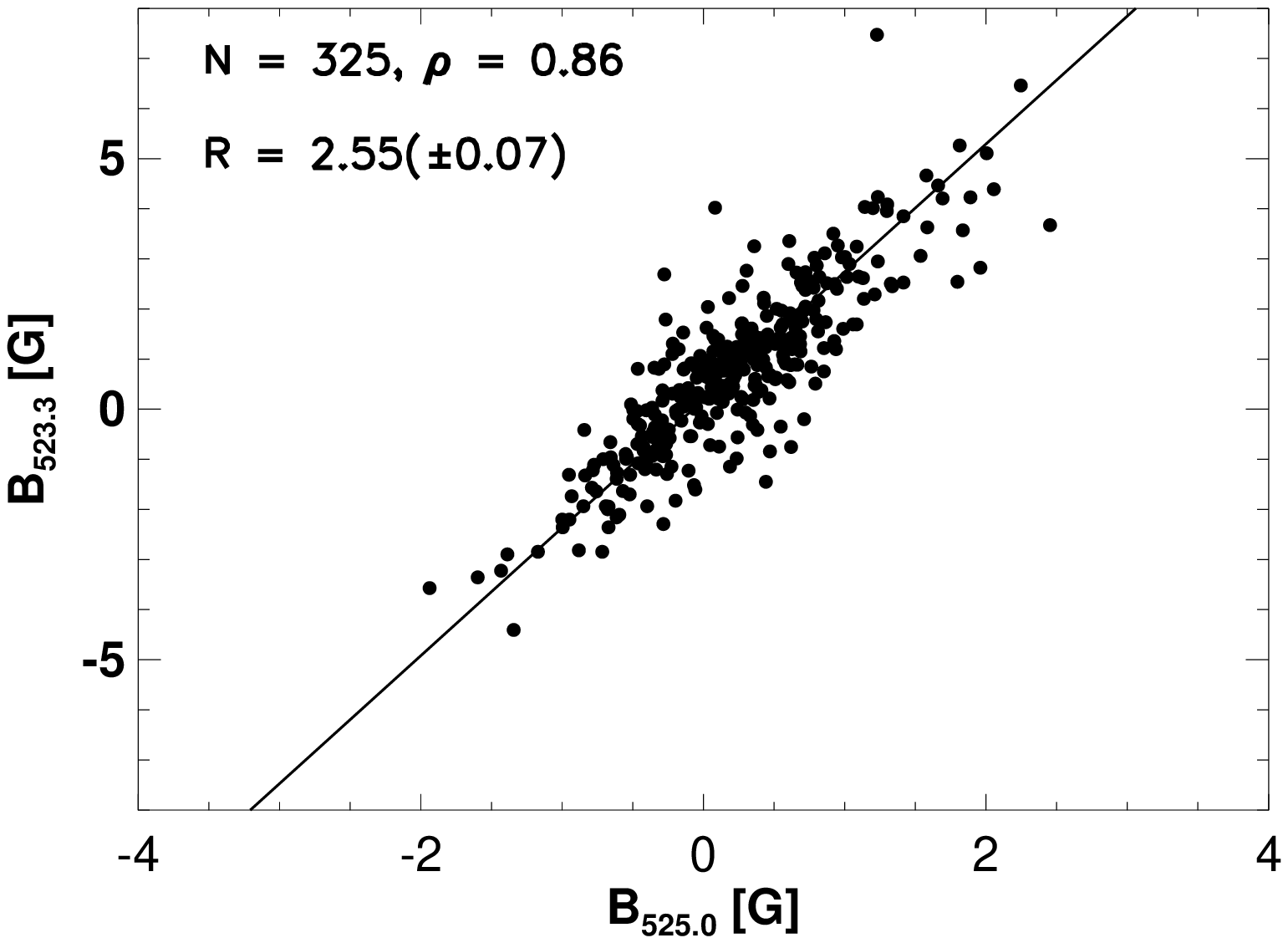}
                             }
     
     \vspace{0.01\textwidth}    

 \caption{Correlation and regression analysis of solar magnetic field observations in 
Fe\,{\sc i}~523.3\,nm and Fe\,{\sc i}~525.0\,nm  for observations with a 
 spatial resolution of $100^{\prime\prime}$, made on 1 February,  2009. Notations are the same as in Figure  \ref{F-scatterplot-U-M}.}          
   \label{F-scat-U-M-100arcsec}
   \end{figure}

Examples of full-disk  magnetograms with a resolution of $100^{\prime\prime}$, obtained on 1 February, 2009 are shown in 
Figure \ref{F-2magnetograms}. Despite of the very weak magnetic field strengths everywhere  across  the disk (minimum of solar activity, there were no sunspots and active regions), both maps show identical spatial structures, and they are well correlated. This is proved by the regression and correlation analysis presented in Figure \ref{F-scat-U-M-100arcsec}. Calculations of the CLV of $R = B(523.3)/B(525.0)$ for these data yield: $R = 2.24 + 0.03\mu + 0.56\mu^{2}$. We see that $R$ is 2.83 at disk center, slightly more than the value of $R = 2.74$ obtained with high-spatial resolution. Despite the decrease of the resolution by a factor of ten, we encounter only a small  $\approx 3\%$ increase of $R$. The difference is much larger for large heliocentric distances, but most probably it is caused by a weakening of the field strengths and by an increasing influence of the noise.

\section{Discussion and Conclusion}
     \label{S-Discussion}

The cornerstone of the study by \inlinecite{Ulrichetal09} is the
difference of magnetic field strengths measured in the different parts of the profile of 
Fe\,{\sc i}~523.3\,nm, \textit{i.e.} the sampling of the line. They wrote: ``contrary to the
assumptions of \inlinecite{FrSt72}, \inlinecite{Stenflo73} and \inlinecite{FrSt78} we find that
the field indicated by $\lambda$\,5233\,\AA{} depends on the spectral sampling'' and that
``we [...] recommend the use of a sampling point as close to the line core as is practical''. 
In the work of \inlinecite{Ulrich92}, the parameters of the spectral configurations with
fiber-optic image reformattors were: centered on 
Fe\,{\sc i}~525.0\,nm -- separation $\Delta =\pm$3.9\,pm; 
centered on Fe\,{\sc i}~523.3\,nm: -- $\Delta = \pm$4.5\,pm. 

In the recent work, \inlinecite{Ulrichetal09} obtained new MWO observations with the same sampling for Fe\,{\sc i}~525.0\,nm and with a new set of separations of $\pm0.9, \pm2.9, \pm8.4,\pm10.2, $and $\pm17.7 \rm{pm}$ for Fe\,{\sc i}~ 523.3\,nm. Computations of magnetic fields from various  spectral sampling pairs and at $\pm8.4\,\rm{pm}$ have shown different values of correlations coefficients $\rho$ and   slopes in the corresponding scatter plots, depending on the value of the separation and the range of $\theta$ (see Figures 1 to 3 and 
Table I in \opencite{Ulrichetal09}). The correlation coefficients for measurements at $\pm8.4\,\rm{pm}$ with the other ones were: 
$\rho = 0.50$ at $\pm0.9\,{\rm pm}$, $\rho = 0.90$ at $\pm2.9\,{\rm pm}$, $\rho = 0.90$ at $\pm10.2\,{\rm pm}$ and $\rho = 0.50$ at $\pm17.7\,{\rm pm}$. 
Only the correlation coefficients for the pairs nearest to 
$\pm8.4\,{\rm pm}$ are reliable. In
two other cases they are too small for a solid statistical analysis.
The examples of slopes obtained by 
\inlinecite{Ulrichetal09} for the central area of the disk (range of $ \sin\theta$  from 0.0 to 0.4) are: 
1.665\,$\pm0.067$ at $\pm0.9\,{\rm pm}$,  1.795\,$\pm0.027$ at $\pm2.9\,{\rm pm}$, 1.162\,$\pm0.010$ at $\pm10.2\,{\rm pm}$ and  2.694\,$\pm0.094$ at $\pm17.7\,{\rm pm}$.

The fact that magnetic fields measured in points near the line center
($\pm2.9\,{\rm pm}$)  are  weaker (by a factor of 0.73 after  application of additional assumptions and  calculations, see Figure 12 of \opencite{Ulrichetal09}) than those
measured in a point near the middle of the line wings ($\pm8.4\,{\rm pm}$) is the basic argument for the decrease of the correction coefficient $R$, following from direct comparison of data sets in Fe\,{\sc i}~523.3\,nm (sampling $\pm8.4\,{\rm pm}$) and Fe\,{\sc i}~525.0\,nm. For the disk center, the value $R = 4.12$ is obtained instead of 5.5.

We are convinced that 
only the Stokes $V/I_{c}$ profile over the whole spectral line can characterize the magnetic field properties in the range of the line formation. But for the comparison with MWO results, 
it is important to find relationships between strengths in the different parts of Fe\,{\sc i}~523.3\,nm in our data.  For this purpose,  calculations with different ``exit slit'' parameters, close to those recently used at MWO, were performed. The results for
3 February, 2009 with  $10^{\prime\prime}$  spatial resolution are given in the following.

The values of ``slit'' widths and  separations are expressed
in the integer numbers of CCD pixels. 
In observations, the width of one  pixel is equal to 0.777\,pm. Following 
\inlinecite{Ulrichetal09}, all data were analyzed relatively to the measurements in 
the part of the wings with a separation of
$\Delta$ = 8.5\,pm and with a ``slit'' width of  $W$ = 9.32\,pm. The comparison of data with these ``slit'' parameters with 
other ones are given in Table~\ref{T-SO}.
For the points at $\pm3.1$\,pm, our result differs from that of \inlinecite{Ulrichetal09}. According to our data, magnetic fields measured closer to the line center are almost the same (only 1.07 times weaker) as measured at $\pm8.5$\,pm. That means, if we correct 
measurements at  
$\pm8.5\,{\rm pm}$ taking into account the regression coefficient $B(\pm8.5\,{\rm pm})/B(\pm3.1\,{\rm pm}) = 1.07$, we get a small difference to measurements in Fe\,{\sc i}~525.0\,nm. Indeed, \inlinecite{Ulrichetal09} found $B(523.3\pm8.4\,\rm{pm}) / B(525.0 \pm3.9\,\rm{pm}) = 5.5$ for the disk center. Applying a special correction factor, they obtained the value of 4.12. In our case we find $5.5 \times (1/1.07) = 5.1$. 

\begin{table}
\caption{Comparison with different ``slit'' parameters.
} 
\label{T-SO}

    \begin{tabular}{rccc}                       
\hline                   
$\Delta\lambda$ [pm]  & Slit width W [pm]  & $\rho$ & $R = B(\pm8.5\rm{pm})/B(\pm\Delta\lambda)$ \\
\hline 
 &&&\\
$\pm$15.53 &  24.82 &  0.991  &  1.271 $\pm$ 0.009 \\
 &&&\\
$\pm$0.8   &  0.80  &  0.870  &  1.018 $\pm$ 0.028 \\
$\pm$3.1   &  1.55  &  0.977  &  1.073 $\pm$ 0.013 \\
$\pm$10.1  &  9.32  &  0.998  &  1.027 $\pm$ 0.003 \\
$\pm$17.8  &  9.32  &  0.935  &  1.528 $\pm$ 0.030 \\  
  &&&\\  
 \hline
\end{tabular}
\end{table}

All results described above have a direct impact on the problem of the calibration of SOHO/MDI magnetic-field data. Some correction coefficients suggested by MWO scientists were already mentioned before. The correction factors for the MDI data are much less if we use the values derived from our observations instead of those obtained by \inlinecite{Tran-etal05} and by \inlinecite{Ulrichetal09}.
For   disk center, we obtain 1.61 /(4.15/2.74) = 1.06 instead of 1.61 according to \inlinecite{Ulrichetal09}.

Our independent estimations of the SOHO/MDI correction factors are obtained using our previous results from the comparison of 
SOHO/MDI data with SSO observations (important: both analyzed data sets correspond to spatial resolution $100^{\prime\prime}$),  published 
by \inlinecite{Demidov-etal08}. From Figure 1(e) of that paper we get an average value of 
2.75 for the B(SOHO/MDI)/B(SSO) ratio. In order to correct the SOHO/MDI data (when they coincide with the corrected data in Fe\,{\sc i}~ 525.0\,nm), we have to multiply them with the coefficient 1.97/2.75 = 0.72, if we take 
$R = B(523.3)/B(525.0)  = 1.97$ from our data with $10^{\prime\prime}$ spatial resolution, and 2.55/2.75 = 0.93, if we take $R = 2.55$ from data (see Figure \ref{F-scat-U-M-100arcsec}) with $100^{\prime\prime}$ spatial resolution. 
Our correction coefficient is smaller by a factor of about two compared to the recent MWO results (\opencite{Tran-etal05}, \opencite{Ulrichetal09}), and instead of increasing the magnetic field strengths from SOHO/MDI 
data we have to decrease them. We would like to note here, that, according to the study of \inlinecite{Demidov-etal08}, observations at Mount Wilson and Sayan observatories made in Fe\,{\sc i}~525.0\,nm are in almost perfect agreement with each other. Therefore, the problem lies in the differences between the observations in the strong line Fe\,{\sc i}~523.3\,nm.

It seems that this result contradicts to the result of \inlinecite{BergerLites03}, 
who obtained a correction factor of
1.56, based on a comparison of SOHO/MDI and {\it Advanced Stokes Polarimeter} (ASP) data. But observations with the ASP were made in a different spectral line (Fe\,{\sc i}~630.25\,nm), and using different lines can lead to quite different results (see \opencite{Gopasyuk73} and \opencite{Demidov-etal08}). Further, \inlinecite{BergerLites03} analyzed only observations made in active regions, instead of full-disk quiet-Sun observations in our case. Besides  that, the observations with ASP and SOHO/MDI were made with much higher spatial resolution.

In connection with this issue it is important to note that 
comparisons of SOHO/MDI with Kitt Peak spectro-polarimeter data (\opencite{JonesCeja01}, \opencite{WenSolKr05}) caused a decrease of 
the SOHO/MDI measurements  by a factor of $\approx1.4$ to adjust them
to the Kitt Peak data.

Summarizing, we arrive at the conclusion that the issue of calibrating SOHO/ MDI data is rather far from  being solved, and new investigations and observations are urgently needed.

\begin{acks}

We appreciate the comments of the unknown reviewer which lead to significant improvements of our paper.
 The results presented in this study were obtained partly under
 support by the DFG (German Science Foundation) grant  BA 1875/5-1.
The MDI data from board the SoHO spacecraft used in this study are
producted by the Stanford-Lockheed Institute for Space Research
and the Solar Oscillations Investigation (SOI) in the W.W.~Hansen
Experimental Physics Laboratory of Stanford University and the
Solar and Astrophysics Laboratory of the Lockheed-Martin Advanced
Technology Center, USA. We took the magnetograms from the web-site
http://soi.stanford.edu/data/ supported by the SOI and MDI team.
This study includes data from the synoptic program at the 150-Foot
Solar Tower of the Mt. Wilson Observatory. The Mt. Wilson 150-Foot
Solar Tower is operated by UCLA, with funding from NASA, ONR and
NSF, under agreement with the Mt. Wilson Institute.

\end{acks}

\end{article}

\begin{thebibliography}{}
 
\bibitem[\protect\citeauthoryear{{Berger} and {Lites}}{2003}]{BergerLites03}
  Berger,~T.E, Lites,~B.W.: 2003, \solphys{} \textbf{213}, 213.

 \bibitem[\protect\citeauthoryear{Davis}{1986}]{Davis86}
Davis,~J.C.: 1986, Statistics and Data Analysis in Geology, John
Wiley, New York.

 \bibitem[\protect\citeauthoryear{{Demidov \etal}}{2002}]{Demidov-etal02}
Demidov,~M.L., Zhigalov,~V.V., Peshcherov,~V.S., Grigoryev,~V.M.:
2002, \solphys {} \textbf{209}, 217.

\bibitem[\protect\citeauthoryear{{Demidov \etal}}{2008}]{Demidov-etal08}
Demidov,~M.L., Golubeva,~E.M., Balthasar,~H., Staude, ~J., Grigoryev,~V.M.:
2008, \solphys {} \textbf{250}, 279.

  \bibitem[\protect\citeauthoryear{{Frazier} and {Stenflo}}{1972}]{FrSt72}
Frazier,~E.N., Stenflo,~J.O.: 1972, \solphys{} \textbf{27}, 330.

\bibitem[\protect\citeauthoryear{{Frazier} and {Stenflo}}{1978}]{FrSt78}
Frazier,~E.N., Stenflo,~J.O.: 1978, \aap{} \textbf{70}, 789.


 \bibitem[\protect\citeauthoryear{Gopasyuk \etal}{1972}]{Gopasyuk73}
Gopasyuk,~S.I., Kotov,~V.A., Severny,~A.B, Tsap,~T.T.: 1973,
\solphys{} \textbf{31}, 307.


\bibitem[\protect\citeauthoryear{{Harvey} and {Livingston}}{1969}]{HarLiv69}
  Harvey,~J., Livingston,~W.: 1969, \solphys{} \textbf{10}, 283.

\bibitem[\protect\citeauthoryear{{Howard} and {Harvey}}{1970}]{HowardHarvey70}
Howard,~R., Harvey, ~J.W.: 1970, \solphys{} \textbf{12}, 23.

\bibitem[\protect\citeauthoryear{{Jones} and {Ceja}}{2001}]{JonesCeja01}
    Jones.~H.P., Ceja,~J.A.: 2001, In Sigwarth,~M. (ed.), 
   \textit {Advanced Solar Polarimetry - Theory,
   Observation, and Instrumentation}, CS\textbf{-236}, Astron. Soc. Pac.,San Francisco, 87.


  \bibitem[\protect\citeauthoryear{{Howard} and {Stenflo}}{1972}]{HSt72}
Howard,~R., Stenflo,~J.O.: 1972, \solphys{} \textbf{22}, 402.

\bibitem[\protect\citeauthoryear{Socas-Navarro \etal}{2008}]{Socas-etal08}
  Socas-Navarro,~H., Borrero,~J.M., Asensio Ramos, A., Collados, M., Dom\'{i}nquez
  Cerde\~{n}a,~I.,Khomenko,~E.V., Mart\'{i}nez Gonz\'{a}lez,~M.J., Mart\'{i}nez Pillet,~V., 
  Ruiz Cobo, B., S\'{a}nchez Almeida,~J.: 2008, \apj{} \textbf{674}, 596.

\bibitem[\protect\citeauthoryear{Solanki}{1993}]{Solanki93}
Solanki,~S.K.: 1993, \ssr{} \textbf{63}, 1.

\bibitem[\protect\citeauthoryear{Stenflo}{1973}]{Stenflo73}
Stenflo,~J.O.: 1973, \solphys{} \textbf{32}, 41.

 \bibitem[\protect\citeauthoryear{Tran \etal}{2005}]{Tran-etal05}
 Tran,~T., Bertello,~L., Ulrich,~R.K., Evans,~S.: 2005, \textit{Astrophys. J. Suppl.}
 \textbf{156}, 295.

 \bibitem[\protect\citeauthoryear{{Ulrich}}{1992}]{Ulrich92}
Ulrich,~R.K.: 1992, In M.S.~Giampapa,~M.S., Bookbinder, ~J.M. (eds.),
Proc. 7th  Cambridge Workshop, \textit{Cool
    Stars, Stellar Systems, and the Sun},CS\textbf{-26}, Astron. Soc. Pac.,San Francisco,265.

 \bibitem[\protect\citeauthoryear{Ulrich \etal}{2002}]{Ul-etal02}
Ulrich,~R.K., Evans,~S., Boyden,~J.E., Webster,~L.: 2002, \textit{
Astrophys. J. Suppl.} \textbf{139}, 259.

  \bibitem[\protect\citeauthoryear{Ulrich \etal}{2009}]{Ulrichetal09}
Ulrich,~R.K., Bertello,~L., Boyden,~J.E., Webster,~L.: 2009, 
\solphys{} \textbf{255}, 53.

\bibitem[\protect\citeauthoryear{{Wang} and {Sheeley}}{1995}]{WangSheeley95}
 Wang,~H, Sheeley,~N.R.,Jr.: 1995, \apj{} \textbf{447},L143.
 
 \bibitem[\protect\citeauthoryear{{Wenzler}, {Solanki},  and {Krivova}}{2005}]{WenSolKr05}
Wenzler,~T., Solanki,~S.K., Krivova,~N.A.: 2005, \aap{}
\textbf{432}, 1057.

 
\textbf{213}, 87.
\end{thebibliography}
\end{document}